\documentclass[showpacs,twocolumn,preprintnumbers,amsmath,amssymb]{revtex4}
\usepackage{amsmath,amsfonts,latexsym,amssymb,graphicx,graphics,epsfig,subfigure,color,makeidx}
\usepackage{xcolor,diagbox}
\usepackage{color}
\usepackage{booktabs}
\usepackage{multirow}
\usepackage{subfigure}
\usepackage[colorlinks,linkcolor=blue,anchorcolor=blue,citecolor=green]{hyperref}
\usepackage{verbatim}

\usepackage{graphicx}
\usepackage{amsmath}
\usepackage{amsthm}
\usepackage{amsfonts,amssymb} 
\usepackage{mathrsfs}

\begin{document}
\title{Hereditary Effects and New Optical Properties of Nonlinear Gravitational Waves}

\author{Yu-Qiang Liu$^{a}$$^{b}$$^{c}$}
\email{yqliu20@lzu.edu.cn}
\author{Yu-Qi Dong$^{a}$$^{c}$}
\email{dongyq2023@lzu.edu.cn}
\author{Yu-Xiao Liu$^{a}$$^{c}$}
\email{liuyx@lzu.edu.cn, corresponding author}

\affiliation{$^{a}$Lanzhou Center for Theoretical Physics,
	Key Laboratory of Theoretical Physics of Gansu Province,
	School of Physical Science and Technology,
	Lanzhou University, Lanzhou 730000, China \\
$^{b}$Department of Basic Courses, Lanzhou Institute of Technology, Lanzhou 730050, China
    \\
	$^{c}$Institute of Theoretical Physics \& Research Center of Gravitation,
	Lanzhou University, Lanzhou 730000, China}

\begin{abstract}
%我们研究了标量张量引力理论中引力非线性微扰的性质。与广义相对论做对比发现额外标量场带来的修改主要体现在奇数阶微扰上。除此之外我们分析了非线性微扰的观测效应。当考虑持续的单色平面波这一特殊情况时，可以证明其振幅本身有一种遗传性效应。这可以使得二阶微扰的振幅的衰减变慢。从而在未来更精确的引力波观测中有可能看到二阶微扰的迹象。
We investigate new optical effects of nonlinear gravitational waves that arise from hereditary effects. Firstly, we show that the amplitude of continuous plane waves has a hereditary effect that grows with distance. This can make the amplitude of nonlinear gravitational waves decay more slowly with distance. This will also lead to observable nonlinear effects in future observations. Secondly, hereditary effects also imply that nonlinear perturbations are nonlocally dependent on linear ones, and this property leads to some special optical effects which allow nonlinear gravitational waves to contain more information. It is even possible to extract information on the distance and direction of the wave source as well as the polarization from the waveform of nonlinear gravitational waves.
\end{abstract}

%\pacs{ 04.50.-h, 11.27.+d}

\maketitle

\section{introduction}
Laser Interferometer Gravitational Wave Observatory (LIGO) directly observed gravitational waves for the first time in 2015 \cite{abbott2016observation}, marked the coming of gravitational wave astronomy era. This event also marks that we can perform direct tests of strong gravitational fields \cite{LIGOScientific:2016lio}. So far there is no solid evidence for the existence of gravity theory beyond general relativity (except for one special signal in the Pulsar Timing Array (PTA) that might imply the existence of non-tensorial gravitational waves \cite{Lee_2008,Chen:2021wdo,Chen:2023uiz}).

Einstein’s general relativity reveals the nonlinear nature of gravity. But since the gravitational interaction is very weak and the wave sources that can be observed are usually very far away from us, it was long assumed that the contribution of the linear-order perturbation was sufficient for gravitational wave observation. This inherent concept was challenged in 1991 by Christodoulou \cite{Christodoulou:1991cr}, who predicted a nonlinear gravitational memory effect whose magnitude can even reach the level of amplitude in linear gravitational waves. Since then, numerous studies have explored the theoretical properties and possible observations of nonlinear memory effects \cite{Blanchet:1992br,Thorne:1992sdb,Wiseman:1991ss,Favata:2010zu,Lasky:2016knh}, as well as nonlinear optics \cite{Harte:2015ila}, etc. and the memory effect has recently been confirmed to be possible to observe in the next galactic core-collapse supernova \cite{PhysRevLett.133.231401}.

Generally speaking, since gravitational nonlinear perturbations decay with distance much faster than linear ones and the wave sources are always far away from us, nonlinear effects are usually ignored in theoretical investigations. However, there is a class of nonlinear effects that can make the amplitude of gravitational waves accumulate over distance, making it possible to produce phenomena strong enough to be observed. These effects are generally referred to as hereditary effects in the literature \cite{Blanchet:1992br,Blanchet:1997ji,Blanchet:2023pce}.

In this Letter, we investigate the observable nonlinear effects of gravitational waves. We show that the amplitude of nonlinear-order perturbations can also accumulate with distance due to hereditary effects. Remarkably, they can even make the amplitude of the second-order gravitational waves decays with distance in the same way as the linear-order ones. This is a distinctive feature of the nonlinear-order perturbations that differs fundamentally from the linear-order ones.

In addition, we also find that nonlinear-order gravitational waves have unique optical effects. For example, nonlinear gravitational waves generated by linear gravitational waves propagate along all spatial directions and have significant interference effects. Besides that, the waveform of the nonlinear-order gravitational waves is also different from the linear one owing to their hereditary effect. These optical effects can help us extract more information from nonlinear gravitational waves, such as the distance and direction of the wave source, as well as the polarization. In other words, for future high precision gravitational wave signals, these effects provide additional parameters that can make the measurement of information, such as the distance and direction of the wave source, more accurate.

In Ref. \cite{Harte:2015ila}, it was shown that a nonindependent breathing polarization mode can also be mimicked for the exact plane wave solution in general relativity. We know that there are only two tensor modes in general relativity but at most six modes in general metric gravity theories \cite{Eardley:1973br,gong2018polarizations,capozziello2006scalar}. And different gravity theories usually have different degrees of freedom of propagation and polarization modes \cite{Maggiore:1999wm,hou2018polarizations,hyun2019exact,Liu:2022cwb,dong2021polarization,Dong:2023xyb,Dong:2022cvf,Dong:2023bgt}. Thus this simulated breathing mode makes it difficult to distinguish different gravity theories only through the polarization properties. In other word, we need more features of gravitational waves to distinguish different gravity theories. The optical effects we study here will provide additional information about the features of gravitational waves.

\section{HEREDITARY EFFECTS AND OPTICAL EFFECTS}
Nonlinear effects of gravitational waves have become more important in recent years. More and more effects of the higher-order perturbations have been proven to be observable. One of the most famous effects is the nonlinear memory effect. However, since the memory effect is non-oscillating, it is difficult to detect it with current ground-based interferometers.

Here we prove that the secondary gravitational waves generated by continue gravitational waves have an oscillating hereditary effect. That is to say, their amplitude can continuously accumulate during the propagation, and the nonlinear gravitational wave at a certain moment should be affected by all the historical retarded solutions before this moment.

\subsection{Asymptotic behavior}

We only focus on the second-order perturbation, since the analysis of higher-order ones shares the same method and leads to similar conclusions. Through the calculation of perturbation of any second-order metric gravity theory, we have the following most general second-order perturbation equation in the flat spacetime background:
 \begin{equation}
 \label{E15} 	\partial_{\lambda}\partial^{\lambda}h_{(2)\mu\nu}=\varepsilon_{\mu\nu}^{\ \ \ \rho \sigma \alpha \beta \gamma \delta}\partial_{\rho}h_{(1)\alpha \beta}\partial_{\sigma}h_{(1)\gamma \delta}+F_{\mu\nu},
 \end{equation}
 where $\varepsilon_{\mu\nu}^{\ \ \ \rho \sigma \alpha \beta \gamma \delta}$ is a product of flat metric factors, the second term on the right hand side comes from the coupling of the gravitational field with extra fields in modified gravity theories and contain no higher than first-order perturbation of metric and second-order perturbation of extra fields. This formula shows that the right hand side of the equation can be regarded as the source of the second-order metric perturbation. Which means that the second-order perturbation is generated by the first-order perturbation in the wave zone. %In other words, the nonlinear-order of gravitational waves are produced by gravitational waves in the wave zone.
 From the perspective of quantum field theory, this is a typical nonlinear feature that gravitons can exchange their own charge.\par
 
 Next, we estimate how the nonlinear gravitational waves decay with distance. Most of the sources we are interested in are very far from us, hence we can treat gravitational waves as plane waves in the wave zone. We assume that there is a continuous monochromatic plane gravitational wave in wave zone. Then, Eq. (\ref{E15}) tells us that every part of this wave has the same effective energy-momentum tensor, and can radiate secondary gravitational waves. So, for an inertial observer at a given moment, these sources distributed in the spacetime can influence the observer’s observation through the retarded solution of gravitational waves. The retarded solution can be written as
 \begin{equation}
 \label{E16}
 	h_{(2)\mu\nu}(x^{\lambda})=\int \frac{1}{|x^{i}-x'^{i}|}T_{\mu\nu}(t-|x^{i}-x'^{i}|,x'^{i})d^{3}x',
 \end{equation}
where $T_{\mu\nu}$ represents all the energy momentum terms in the right hand side of Eq. (\ref{E15})
\begin{equation}
	T_{\mu\nu}(t-|x^{i}-x'^{i}|,x'^{i})=\varepsilon_{\mu\nu}^{\ \ \ \rho \sigma \alpha \beta \gamma \delta}\partial_{\rho}h_{(1)\alpha \beta}\partial_{\sigma}h_{(1)\gamma \delta}+F_{\mu\nu}.
\end{equation}
\begin{figure}
		\centering
\includegraphics[width=10cm]{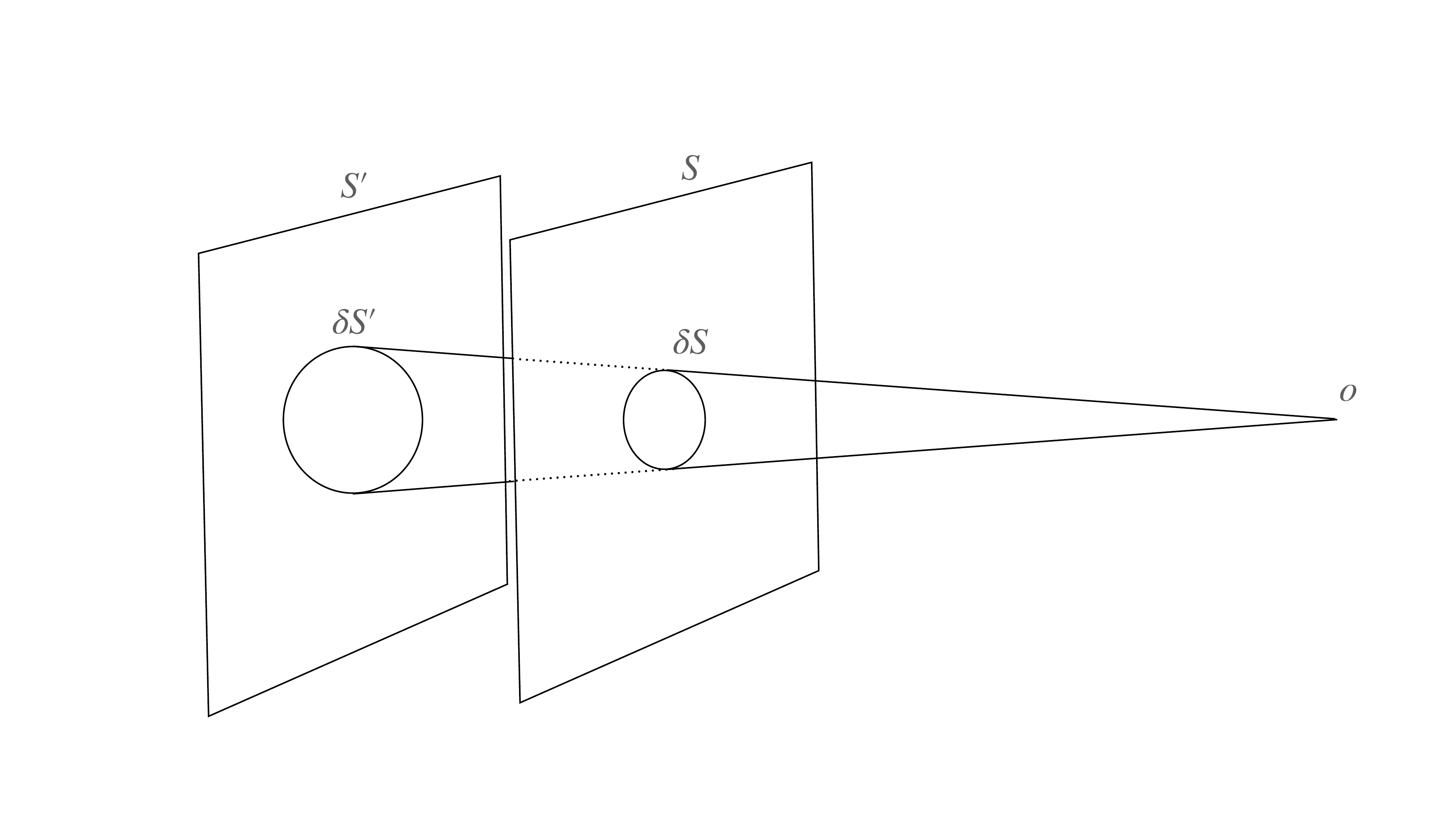}
\caption{
Assuming a continuous plane wave propagating from left to right, $S$ and $S'$ are two wavefronts, $\delta S$ and $\delta S'$ are two neighboring regions, their energy will generate second-order gravitational waves that affect the perturbation amplitude at o through Eq. (\ref{E16}).}
\label{p1}
\end{figure}

Let us consider the schematic plotted in FIG. \ref{p1}, where we have a monochromatic plane gravitational wave propagating continuously from left to right, we define this propagation direction as $+z$. Planes $S$ and $S'$ represent two different wavefronts, and $o$ is the location of the observer.

Now, we analyze the effect of the second-order gravitational waves radiated by the two wavefronts. As shown in FIG. \ref{p1}, the gravitational wave in the infinitesimal region $\delta S$ is responsible for this effect of the gravitational waves propagating along $+z$ at $o$. Since the decay with distance of gravitational wave energy follows the inverse square law, and the area of $\delta S$ increases proportionally with the square of the distance. Therefore, the gravitational waves generated by any two different wavefronts (e.g. $\delta S$ and $\delta S'$) have the same effect on the observer.

Therefore, the second-order gravitational waves observed by the observer depend nonlocally on all historical retarded solutions. From Eq. (\ref{E15}), it can be seen that the second-order perturbation contains at most the square term of the first-order perturbation, its amplitude and its distance from the wave source satisfy the following relationship
\begin{equation}
	h_{(2)}\propto \frac{1}{r^{2}}.
\end{equation}

Then, after considering the hereditary effect of amplitude,
\begin{equation}
	\begin{split}
		h_{(2)\mu\nu}(x^{\lambda})&=\int \frac{1}{|x^{i}-x'^{i}|}\mathscr{T}_{\mu\nu}(t-|x^{i}-x'^{i}|,x'^{i})d^{3}x',\\
		&=\int \mathscr{T}_{\mu\nu}(t-|z-z'|,z')d^{3}z'
	\end{split}
\end{equation}
where the second equal sign means that because each wavefront has the same contribution to the observer, after completing the integration in the $x-y$ plane, the integration in the $z$ direction is the superposition of all wavefronts with the same contribution. This means the intensity of the secondary gravitational waves at the observer's location is the sum of the contributions of the secondary gravitational waves emitted by all sources at $x'$.

Therefore, after taking into account the hereditary effect of amplitude, the asymptotic behavior of the second-order gravitational waves at a distance is
\begin{equation}
	h_{(2)|\text{hereditary}}\propto \frac{1}{r}.
\end{equation}
It is worth noting that the asymptotic behavior of $h_{(2)|\text{hereditary}}$ is the same as the linear-order one for continue plane waves.

But if we want to link the above results with actual observations, we need to treat the spherical waves. Since most of the sources we are interested in are far away from us, we can always treat gravitational waves as plane waves near the observer. For example, if we can treat the gravitational waves as plane waves when the distance between the gravitational wave and the observer is less than $\frac{r}{10^{3}}$. Thus, the asymptotic behavior of $h_{(2)|\text{hereditary}}$ far away from the source is still
\begin{equation}
	h_{(2)|\text{hereditary}}\propto \frac{r}{10^{3}}\times \frac{1}{r^{2}}\propto \frac{1}{r}.
\end{equation}
This will make it possible to directly observe nonlinear gravitational waves in more precise gravitational wave detection in the future, allowing further insights into the nature of gravity.

\subsection{Optical properties}
It is well known that the gravitational radiation is emitted in all directions, hence the secondary radiation from the regions outside the neighboring regions $\delta S$ in FIG. \ref{p1} can also affect the observer at point $o$. We find that gravitational waves propagating from different directions to the observer can produce special interference phenomena.

Although we demonstrated in the previous section that the secondary gravitational waves generated by $\delta S$ and $\delta S'$ in FIG. \ref{p1} contribute the same amplitude to the ob- server, $\delta S$ and $\delta S'$ in different propagation directions usually have different phases. This is because the distances between $\delta S$ and $\delta S'$ are different in different directions, so the superposition of these waves will multiply or cancel each other. After simple calculation, the angular distribution of the amplitude of gravitational waves can be obtained as
\begin{equation}
	I=\left|\cos \big(2\pi(\sec \theta-1)\big)\right|
\end{equation}

\begin{figure}
		\centering
\includegraphics[width=6cm]{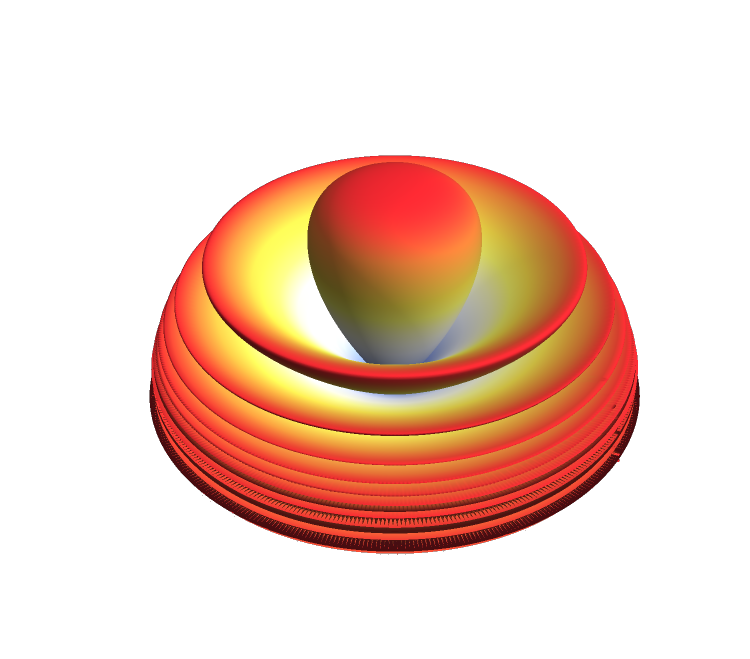}
\caption{
Schematic diagram of the amplitude distribution of nonlinear gravitational waves at the observer’s location. The $z$-axis is vertically upwards, and $\theta\in[0,\pi/2]$. The distance between each point on the surface and the origin represents the relative amplitude of the nonlinear gravitational wave in the corresponding direction.}
\label{p2}
\end{figure}

Figure \ref{p2} is a schematic diagram of this interference intensity. The vertical direction in the figure represents the $z$-direction, and the distance from any point on the two-dimensional surface to the origin represents the relative amplitude of gravitational waves in the corresponding direction. That is, $h_{(2)|\text{hereditary}}$ has interference effects at different directions, which is fundamentally different from linear gravitational waves.

\subsection{Observation properties}
Now, we have analyzed the asymptotic behavior and new optical properties of nonlinear gravitational waves. The asymptotic behavior ensures the secondary gravitational waves could be observable. So, the next step is to analysis the observation properties of these optical effects and how these properties feed back into the theoretical research on gravity.

Firstly, the interference intensity distribution in FIG. \ref{p2} is not the intensity distribution of the nonlinear gravitational waves that we truly detected. It is also necessary to consider the angular dependence of the radiation intensity of gravitational waves. In other words, the amplitudes of gravitational waves in different direction emitted from the same wave source are generally different. In order to obtain the spatial angular distribution of gravitational radiation with different polarization modes, we need to consider the density of gravitational radiation energy flow:
\begin{equation}
	F=\frac{1}{32\pi}\omega^{2}\langle h_{\mu\nu}h^{\mu\nu}\rangle.
\end{equation}

\begin{figure*}
\centering
\subfigure[Tensor mode]{
\includegraphics[width=4cm]{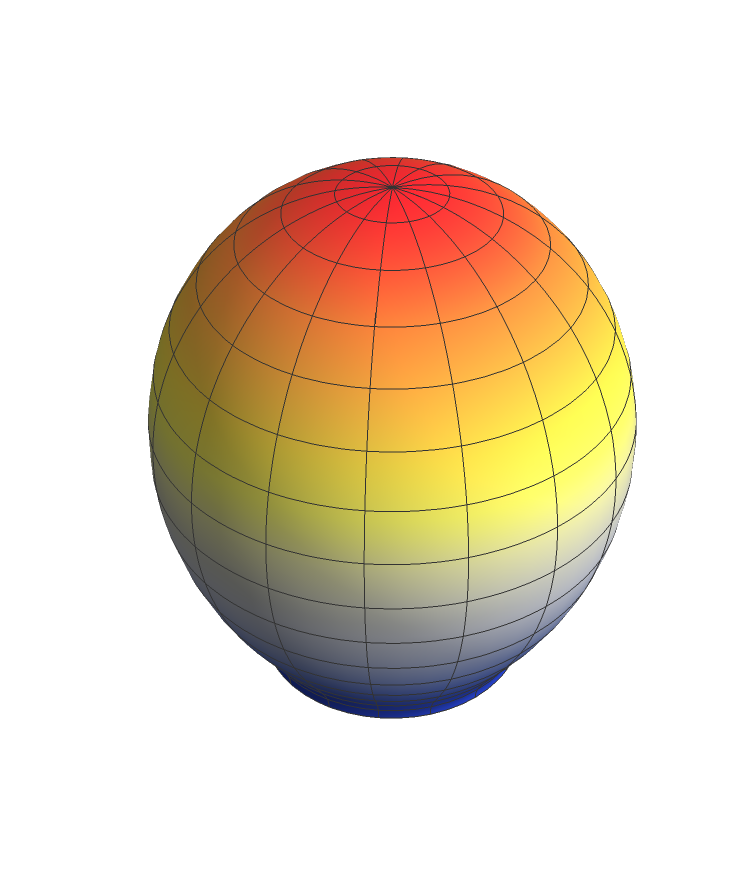}}
\quad
\subfigure[Breathing mode]{
\includegraphics[width=4cm]{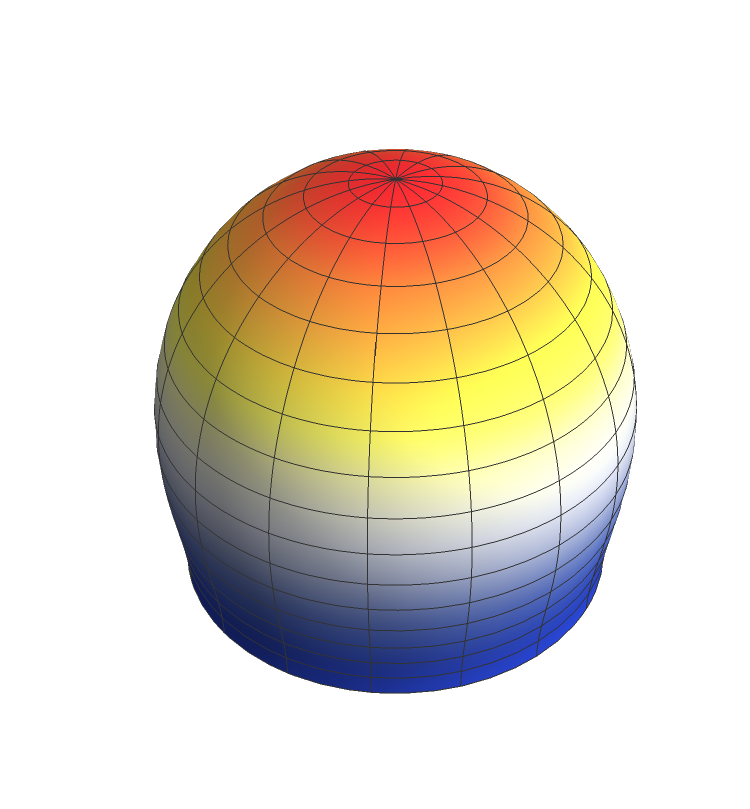}}
\caption{The angular distribution of amplitude of gravitational waves. The $z$-axis is vertically upwards, and $\theta\in[0,\pi/2]$.}
\label{p3}
\end{figure*}

\begin{comment}

\begin{figure*}[htbp]
\centering
\begin{subfigure}[t]{0.45\textwidth}
\centering
\includegraphics[width=5cm]{tensor_modes.pdf}
\caption{Tensor mode}
\label{fig:subfig1}
\end{subfigure}
\hspace{0.01\textwidth} % 调整两张图片之间的间距
\begin{subfigure}[t]{0.45\textwidth}
\centering
\includegraphics[width=5cm]{breathing_modes.pdf}
\caption{Breathing mode}
\label{fig:subfig2}
\end{subfigure}
\caption{The angular distribution of amplitude of gravitational waves. The $z$-axis is vertically upwards, and $\theta\in[0,\pi/2]$.}
\label{p3}
\end{figure*}
\end{comment}

The density of gravitational radiation energy flow can be calculated by substituting the amplitude of gravitational waves calculated in the previous subsection into the above equation. The angular distributions of the amplitude of tensor mode and breathing mode are shown in FIG. \ref{p3}. It can be seen that for transverse gravitational waves, the radiation energy is highest along the $z$-axis direction and lowest perpendicular to the $z$-axis direction. And the angular distributions of different polarization modes are generally not the same.

Therefore, by combining the angular distribution of interference effect in FIG. \ref{p2} and the angular distribution of radiation intensity of wave source in FIG. \ref{p3}, the angular distribution of the intensity of the non-linear gravitational wave can be obtained at the observer.

In addition, we can detect secondary gravitational waves in multiple directions, as shows in FIG. \ref{p2}. Therefore if we can distinguish these gravitational waves in different directions, we can obtain the direction of the wave source.

Secondly, the waveform of a nonlinear gravitational wave is also different from that of a linear one. This is mainly because the nonlinear perturbation of gravity depends nonlocally on the energy distribution of gravitational waves in whole spacetime, which is exactly a typical hereditary effect.

\begin{figure}
		\centering
\includegraphics[width=5cm]{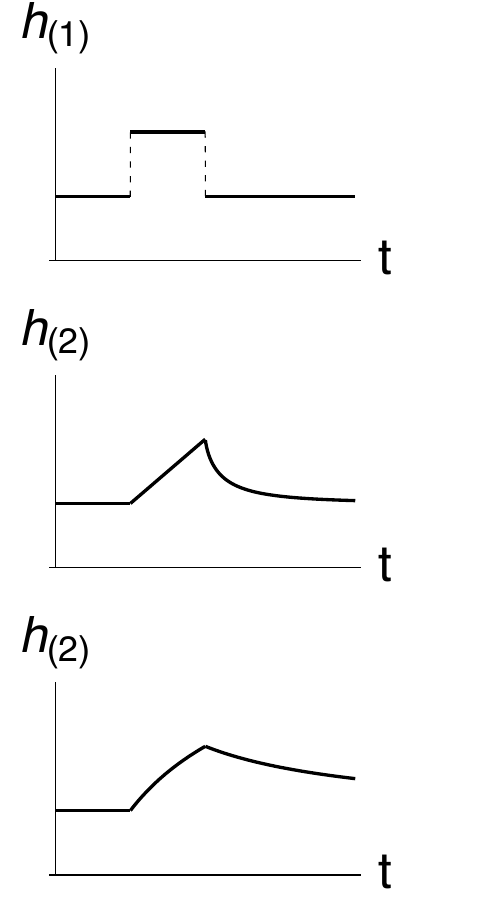}
\caption{
The corresponding change in the amplitude of the nonlinear harmonic gravitational wave when the amplitude of the linear harmonic gravitational wave changes. The duration of gravitational waves is $\Delta t$. The second and third figures show the amplitude changes of the second-order gravitational waves received by different observers at distances from wave sources $c \Delta t$ and $10 \times c \Delta t$.}
\label{p4}
\end{figure}

Here we take a simple example to show this difference. In FIG. \ref{p4}, the amplitude of a linear harmonic gravitational wave suddenly increases at a certain moment and then recovers after a period of time $\Delta t$. The second and third figures in FIG. \ref{p4} show the waveforms of the second- order harmonic gravitational waves observed at different distances from the wave source. To calculate the amplitude values at each moment, we need to accumulate all secondary gravitational waves that can affect the observer at this moment. That is, the amplitude of each point in the waveform of $h_{(2)|\text{hereditary}}$ needs to be integrated throughout the entire historical spacetime.

It is worth noting that, as shown in FIG. \ref{p4}, the amplitude of the nonlinear gravitational wave also has a tail effect. And its waveform depends on the distance between the wave source and the observer, as shown in the second and third figures in FIG. \ref{p4}. In other words, the waveform of nonlinear gravitational waves contains distance information. The final stage of the $h_{(2)|\text{hereditary}}$ is a slow decay stage, because the current nonlinear gravitational wave also depends on the linear gravitational wave at any time in the past.

Therefore, due to the nonlocal dependence of the entire history of nonlinear gravitational waves, nonlinear gravitational waves will provide us with more information about polarization modes, wave source direction, and distance.

\section{conclusion}
In this letter, we explored optical properties and the possible observational effects of the higher-order gravitational waves. Firstly, for the continuous plane gravitational waves, the amplitude of the higher-order perturbations has a hereditary effect that can make itself accumulate with distance. This can make the decay of the higher-order perturbation amplitude slower. For the second-order gravitational perturbation, it can even make its amplitude decay with distance in the same way as the linear-order ones. Therefore, contrary to intuition, it is possible to detect nonlinear effects in more accurate observations in the future.

If these nonlinear effects are observed, it is possible to discover their unique optical effects provided in the last section. The direction and distance of the wave source and the polarization modes of gravitational waves can be reflected in the angular distribution and waveform of the nonlinear gravitational waves. Therefore, these important parameters of gravitational waves and wave sources will be measured more accurately when consider these nonlinear effects.

  \section*{Acknowledgments}
  This work is supported in part by the National Key Research and Development Program of China (Grant No. 2021YFC2203003), the National Natural Science Foundation of China (No. 12247101 and No. 123B2074), the 111 Project under (Grant No. B20063), the Major Science and Technology Projects of Gansu Province, and Lanzhou City's scientific research funding subsidy to Lanzhou University, the Department of education of Gansu Province: Outstanding Graduate ``Innovation Star" Project (Grant No. 2023CXZX-057).
  
%This work is supported in part by the National Key Research and Development Program of China (Grant No. 2020YFC2201503), the National Natural Science Foundation of China (Grants No. 11875151 and No. 12047501), the 111 Project (Grant No. B20063), the Department of education of Gansu Province: Outstanding Graduate ``Innovation Star" Project (Grant No. 2023CXZX-057) and Lanzhou City's scientific research funding subsidy to Lanzhou University.
%\appendix
%\section{123}\label{appendix:1}

\bibliographystyle{unsrt}
\bibliography{reference2}

\end{document}